*Review*

# Drugs and drug delivery systems targeting amyloid-β in Alzheimer's disease

**Morgan Robinson [1], Brenda Yasie Lee [1] and Zoya Leonenko [1, 2, 3,] ***

[1] Department of Biology, University of Waterloo, Waterloo, Ontario, Canada
[2] Department of Physics & Astronomy, University of Waterloo, Waterloo, Ontario, Canada
[3] Waterloo Institute for Nanotechnology, Waterloo, Ontario, Canada

**\* Correspondence:** Email: zleonenk@uwaterloo.ca; Tel: +1 519-888-4567 ext. 38273.

**Abstract:** Alzheimer's disease (AD) is a devastating neurodegenerative disorder with no cure and limited treatment solutions that are unable to target any of the suspected causes. Increasing evidence suggests that one of the causes of neurodegeneration is the overproduction of amyloid beta (Aβ) and the inability of Aβ peptides to be cleared from the brain, resulting in self-aggregation to form toxic oligomers, fibrils and plaques. One of the potential treatment options is to target Aβ and prevent self-aggregation to allow for a natural clearing of the brain. In this paper, we review the drugs and drug delivery systems that target Aβ in relation to Alzheimer's disease. Many attempts have been made to use anti-Aβ targeting molecules capable of targeting Aβ (with much success in vitro and in vivo animal models), but the major obstacle to this technique is the challenge posed by the blood brain barrier (BBB). This highly selective barrier protects the brain from toxic molecules and pathogens and prevents the delivery of most drugs. Therefore novel Aβ aggregation inhibitor drugs will require well thought-out drug delivery systems to deliver sufficient concentrations to the brain.

**Keywords:** amyloid; amyloid aggregation; inhibitor drugs; Alzheimer's disease; drug delivery systems; blood brain barrier; nanotechnology

**Abbreviations:** Alzheimer's disease (AD), amyloid-beta (Aβ), blood-brain barrier (BBB), amyloid precursor protein (APP), polyethylene glycol (PEG), monoclonal antibodies (MAb), beta-secretase 1 (BACE1), passive immunotherapy (PI), atomic force microscopy (AFM), retro-inverso (RI), molecular dynamics (MD), cell-penetrating peptide (CPP), polylactic-co-glycolic acid (PLGA), nanoparticle (NP), tight junctions (TJ), transendothelial electrical resistance (TEER), P-glycoprotein (P-gp), advanced glycation end products (RAGE), transferrin receptor (TfR), diphtheria toxin receptor (DTR), receptor-mediated transcytosis (RMT), low density lipoprotein related protein (LRP1), apolipoprotein E (ApoE) rabies virus glycoprotein (RVG), gold nanoparticle (AuNP), poly lactic acid (PLA), absorptive-mediated transcytosis (AMT), transactivator of transcription (TAT), Madin-Darby canine kidney (MDCK)

## 1. Introduction

With people living longer than ever and aging populations increasing around the world, the health concerns associated with age will continue to increase. Those afflicted with Alzheimer's disease represent a very large portion of the geriatric population. Accounting for 50–80% of cases of dementia, AD is the leading cause of dementia worldwide, with over 36 million registered cases in 2009. It is expected that the incidence of AD will double every twenty years, with approximately 66 million worldwide by 2030 and 115 million cases globally by 2050 [1]. With no cure available, specific treatment options for AD are limited to treating the symptoms of dementia, not the root cause. Five FDA approved drugs are available for improving memory and cognitive function in AD patients. These drugs attempt to increase the amount of neurotransmitters in the brain, which have been shown to improve cognition in mild to moderate AD patients for 6 months [2]. Not only are the treatment methods non-specific to AD but a sensitive and specific diagnostic test for AD is still unavailable. The only conclusive diagnosis for AD is performed during a postmortem autopsy on the brain, where plaques can be directly visualized.

Early detection of AD is near but not yet reliable and preventative treatments in the pre-dementia phase (called prodromal AD) are not available as it has only recently been defined. Psychiatric evaluations involving memory tests are the standard diagnostic tool with neural imaging used as a supplemental diagnostic and is a new and exciting emerging technology which will be useful for diagnosing prodromal AD [3,4]. Lack of detection is especially concerning as evidence suggests symptoms may require a decade to manifest after the onset of pathophysiology in AD [5].

*The amyloid beta (Aβ) hypothesis*

There are several theories which have been proposed to explain the progression of Alzheimer's disease with the amyloid cascade hypothesis remaining the most widely accepted. This hypothesis states that the primary neurologic insult is caused by toxic and soluble Aβ oligomers. The trans-membrane amyloid precursor protein (APP) is cleaved by β and γ secretases to produce Aβ monomers [4], which misfold to form toxic β-sheet oligomers and eventually form larger fibrils and plaques. After the enzymatic processing of APP, Aβ monomers are free to misfold back onto themselves through side chain interactions producing a hairpin structure with residues in the central region and N-terminus forming intermolecular hydrogen bonds with other monomers [6]. Binding sites for monomers always appear on the outside of the growing oligomer, turning into nucleation sites for further growth into toxic oligomers and eventually fibrils and plaques.

Literature in the last decade has implicated small oligomers as the most toxic of amyloid aggregate species, with plaques and fibrils being non-toxic even though they may still be a source of free amyloid [7]. Current reports suggest that non-specific interactions of toxic soluble amyloid oligomers with the cell membrane play an important role in the mechanism of cytotoxicity [7,8].

Recent studies in amyloid lipid membrane interactions have been able to show that amyloid oligomers aggregate on the surface of the cell membrane, deforming it [7-9]. In extreme cases, trans-membrane channels [10-12] and pores are formed that disrupt membrane integrity [13,14]. Effects on the cell membrane caused by amyloid may be more subtle, including: alterations in synaptic plasticity, receptor protein distribution and modification of signaling pathways prior to the occurrence of more severe deformations [8,13]. Aβ molecules are found in all humans, regardless of age or disease but the natural role of these amyloid are largely unknown. It has been suggested that the Aβ monomer may play a role in important signaling pathways in the brain [15] and is likely to have neuroprotective properties at low concentrations [16]. It appears that healthy individuals are not susceptible to amyloid induced cytotoxicity and can clear amyloid from the brain before it reaches neurotoxic levels by balancing amyloid production and clearance [17]. If amyloid burden can be reduced, it is speculated that it may be possible to slow the progression of Alzheimer's disease.

## 2. Therapeutic strategies in AD targeting amyloid aggregation pathways

Insight into amyloid aggregation has led to various approaches in the last two decades to slow or prevent amyloid aggregation and improve clearance from the brain. One approach is to prevent aggregation by using molecules/ligands which directly bind to and modify or inhibit aggregation of amyloid. These include: PEG nanoparticles [18], lipid based nanoparticles containing phosphatidic acid and cardiolipin [19], small molecules (like curcumin, melatonin) [20-24], monoclonal antibodies (MAbs) directed against Aβ [25-30] and various other peptidic aggregation inhibitors [31-35]. Another approach that has been proposed with favorable results is active immunotherapy or Aβ vaccines; this was found to lessen amyloid burden on the brain in mouse models [36]. However, concerns with safety have limited the use of active amyloid vaccines. In one clinical trial, 6% of patients developed meningoencephalitis, a dangerous inflammation of the meninges and the brain [37]. Antibodies against amyloid, synthetic or native, signal microglial cells to facilitate uptake and autophagy of amyloid, clearing it from the brain. It is also possible to target other clearance mechanisms, upregulating proteins responsible for moving amyloid out of the brain, via receptor mediated BBB efflux or enzymatic degradation pathways [38]. A third approach is to prevent aggregation at the source by targeting its precursor, either APP itself or the enzymes which cleave it. The effects of APP and secretase inhibition is not well established and dangerous consequences on downstream pathways are suspected as they play roles in many neural processes [39,40].

### 2.1. APP cleavage by secretases

Reduction in amyloid burden can be achieved by targeting the enzymatic pathway responsible for cleaving APP into Aβ monomers. The rate limiting enzymatic step involves beta-secretase 1 (BACE1) cleavage of APP and presents a key target for inhibition [39]. Secretase inhibitor strategies include RNA interference to prevent translation of the enzyme, small molecules to reduce the enzyme's activity and MAb to clear the enzyme [41-44]. Lentiviral vectors expressing siRNAs directed against BACE1 transcripts were shown to reduce amyloid production and cognitive deficit in APP transgenic mouse models [43]. Small molecule inhibitors of secretases are non-specific and have displayed unpredictable off target effects that have led to cancellation of some clinical trials [44], while larger more specific therapeutics show limited BBB permeation [41]. Antibodies to BACE1 were discovered to specifically target BACE1 and prevent amyloidosis in human cell lines, primary neurons, and *in vivo* mouse and non-human primate models, but BBB permeation still represents a large barrier to their effectiveness [41]. Although the knockout of BACE1 in mouse studies yielded insignificant consequences [45], the effects on BACE1 substrates (such as myelination, retinal homeostasis, synaptic function and brain circuitry) are still not well understood and more studies are required. Combination therapies represent an exciting opportunity to achieve improved amyloid reduction in the brain but have not been explored in great detail. Recently, anti-Aβ monoclonal antibody Gantenerumab directed at amyloid and a small molecule BACE1 inhibitor were assessed in a London mouse model of AD [46]. Modulators and inhibitors of gamma secretases have also been explored, complete inhibition is not feasible due to the role it plays in critical Notch signaling pathways, while modulators are able to preferential target gamma secretase activity on APP alone and are of great therapeutic interest [40]. With at least 10 small molecule BACE1 inhibitors in current clinical trials, BACE1 represents a key target for therapeutic intervention of amyloid targeting intervention.

## 2.2. Passive immunotherapy targeting amyloid

Passive Immunotherapy (PI) uses MAbs with high-specificity to the various species of Aβ to reduce aggregation and promote immunogenic clearance from the brain [28,29]. Three non-mutually exclusive pathways for the PI mechanism have been suggested and may all act in parallel [47]. MAbs are expected to directly trigger an immune response against Aβ deposits in the brain, increasing microglial uptake [29]. MAbs can also prevent aggregation of and disaggregate fibrils and plaques through competitive processes; this is supported by early AFM studies on MAbs like m266.2 [27]. MAbs have been suggested to improve cognition and amyloid burden in mouse models through a proposed "sink" effect where the reduction of free soluble amyloid in the blood causes a shift in amyloid equilibrium to favor efflux from the brain [26]. MAbs do not cross the BBB efficiently, reaching a maximum of 0.11% at 1 hour after injection, which adds weight to the sink hypothesis [48]. In AD, BBB integrity is compromised and therefore increases the risks associated with over-activating microglial cells and pro-inflammatory response. Various dangerous risks including cereberal amyloid angiopathy and other conditions have been reported [49-52]. Attempts to engineer safer MAbs, with reduced effector response and improved BBB permeation, are currently being explored [29]. Another obstacle in using PI is that while largely successful in many preclinical mouse models, both at reducing amyloid burden and improving cognition [25,26,30,52], MAb therapies for use in humans has been unsuccessful at improving the key markers of AD (cognition, memory and learning) in clinical trials [53-56].

Solanezumab (Sol) is the humanized IgG1 MAb m266.2, which recognizes the central region of Aβ$_{13-28}$ and binds soluble amyloid species. Solanezumab was shown to have no significant effect on improving the primary outcomes of patients in two phase 3 clinical trials, though no adverse events could be associated with it [53]. A more detailed analysis yielded mildly encouraging results for a subgroup of mild AD patients [53]. This trial is ongoing while other preventative trials of Solanezumab have begun. Bapineuzumab (which recognizes N-terminal amyloid and binds to fibrils and plaques) was shown to have no significant effect on improving cognition of patients in two phase 3 clinical trials. Coupled with a safety concern, where MRI imaging showed abnormalities associated with vasogenic edema, clinical trials were discontinued [55]. These setbacks have led experts to a shift in expectations for amyloid intervention therapies and suggest that PI may not be effective after the onset of symptoms due to the significant amount neuron loss [54,56]. To test this premise, clinical trials have shifted to preventative studies, with the newcomer Crenezumab launching a five-year preventative trial in 2012 [57]. Crenezumab was engineered with an IGg4 backbone which causes a milder immune response yet still binds with high affinity to amyloid monomers, oligomers and fibrils; it was shown to be neuroprotective and increased uptake of Aβ by microglial cells in mouse models [25,57]. Another PI called Gantenerumab, which binds both the *N*-terminus and mid-region of Aβ but predominantly fibrillar species, was being tested in phase 3 trials with prodromal (pre-dementia) and mild AD patients [58]. However these studies have been cancelled due to a lack of efficacy, findings from the trial have not yet been published. Since amyloid insult may occur as early as 20 years prior to the onset of symptoms [5], early diagnostic testing stands as an important obstacle to effective treatments. As paradigms shift to preventative strategies, engineered MAbs and other peptide inhibitors of amyloidosis represent potentially safer and more effective alternatives.

### 2.3. Peptide inhibitors of amyloid aggregation

In 1996, Tjernberg et al. reported the use of amyloid peptide fragment KLVFF as an aggregation inhibitor and showed proof of principle for the use of peptide-based ligands built from the sequence of amyloid itself [35]. Although aggregation is still seen, a noticeable decrease in fibrillization suggests that these peptides may serve as ligands with the ability to affect the dynamics of fibril formation. This study systematically tested 31 decamer Aβ fragments, selected the decapeptides with highest affinity, truncated and mutated them to determine the minimum sequence needed to prevent binding, and finally identified the pentapeptide KLVFF (Aβ$_{16-20}$) [35]. They found that the sequence KLXXF was critical for amyloid binding. The same year, Ganta et al. showed how Aβ$_{15-25}$ attached to a disrupter element (repeated oligolysine) could reduce amyloid beta toxicity. Although this peptide inhibitor did not block beta sheet interactions and fibril formation, it did cause changes in aggregation kinetics and higher order structural changes in fibrils by shortening the length of the fibrils while increasing the amount of fibril entanglement [32]. This highlights the importance of how the aggregated structure of Aβ can affect cytotoxicity. Pallitto et al. modified the recognition sequence KLVFF by adding repeating oligoproline units. The ring structure of the proline side chain disrupts beta sheet interactions, limiting the stacking ability of amyloid to grow into larger fibrils through a dynamic competitive process [34]. They also found that shuffling the KLVFF amino acid sequence still inhibited Aβ fibril formation and had nearly identical binding characteristics. This implies that the overall hydrophobicity of the amyloid ligand is important for efficient binding with amyloid [34]. These early preliminary studies have paved the way for the advanced design of next generation peptide inhibitors.

With a foundation of knowledge built from early amyloid studies, it is possible to design effective amyloid aggregation inhibiting drugs with various compositions for stability, binding affinity, cell membrane and BBB permeability, and immune system evasion. Small peptides under nine amino acids made with synthetic amino acid residues—N-methylated, dextrorotary (D)—improves immune system evasion, proteolytic stability, and also interactions with the target amyloid [59,60]. A peptide inhibitor, called OR2, was designed from the KLVFF sequence, modified with a glycine spacer and charged amino acid residue at each terminus. These charged residues improve aqueous solubility and disrupt fibril formation. OR2 was shown to modify early aggregation of Aβ and protect SHSY-5Y cells from Aβ cytotoxicity [61]. Later, to improve proteolytic stability and reduce immune response they substituted various amino acids with their corresponding D-enantiomer. In order to maintain the biological activity of the original peptide a simple swap will not due, reversal of the peptide bond is also required. This "*retro inverso*" version of the peptide, newly designated RI-OR2, was shown to be effective at inhibiting oligomerization and improving the survival of SH-SY5Y cells against Aβ toxicity, while also remaining stable in human blood serum and brain extract for at least 24 hrs [62].

The design and screening of potential drugs and protein therapeutics which bind to and prevent oligomerization using Computer Aided Drug Design (CADD) is a rapid and cost-effective technique for developing and screening drug candidates [63]. Procedurally, the amyloid oligomerization inhibitor drug design begins with a basic lipophilic amyloid recognition sequence: KLVFF, LVFFAE and KVLFFAE, as identified in earlier studies [35]. Several amino acids in the sequence are *N*-methylated to help prevent the inhibitor from contributing to the growth of Aβ oligomers, also *N*-methylation may improve membrane permeability [64]. Next, various modifications and substitutions to the peptide are made, such as additions of γ-diaminobutyric acid as an *N*-terminal residue for improved interactions with amino acid $D_{23}$ and substitution of lysine with ornithine, a synthetic amino acid which improves electrostatic side chain interactions with $E_{22}$ [63]. Other improvements to peptide inhibitors can be made by substituting various lipophilic residues, which may optimize hydrophobic interactions between the drug and amyloid target. Alternatively, one can substitute lipophilic aromatic amino acids which have been shown to be important for peptide and protein recognition including amyloid [60]. With this series of inhibitors (labeled the SG series), leading candidates based on MD simulations were tested for their ability to block aggregation using thioflavin T fluorescence assay, western blot and circular dichroism [63]. Strong correlations between success in MD simulation and thioflavin T fluorescence assays were found, highlighting the success of in silica drug design and screening. Furthermore, it has been confirmed in our lab group using atomic force spectroscopy that SG inhibitors are able to reduce binding events between single amyloid monomers [33]. Currently, various SG inhibitors with different modifications, D-amino acid incorporation and various predicted binding orientations are being screened and tested both through direct force measurements and *in vitro* assays.

Peptide based inhibitors present an alternative and appealing preventative strategy to MAb therapies, as they are not as costly to produce, are smaller in size, versatile and intrinsically safer. Antibody fragments and Fc engineered MAbs, which are designed to be safer alternatives to traditional PI, do not offer significant advantages over peptide based inhibitors. Peptide inhibitors are also easily modified for superior BBB permeation through the addition of targeting ligands and shuttling molecules. Parthsarathy et al. modified a peptide inhibitor with a cell-penetrating peptide (CPP) derived from an HIV regulatory protein. This was shown to improve delivery of the peptide to cells and the brain, showing improvement in a transgenic mouse model [65]. Through a more complicated process, MAbs can also be improved by making them bi-specific, able to bind amyloid and some feature of the BBB for improved brain delivery [29,66]. Immunogenic clearance of Aβ has been the focus of research, however as paradigms in AD treatment shift to prevention, inhibition of Aβ aggregation may be suitable to allow natural clearing, to which peptide inhibitors have the advantage.

*2.4. Nanoparticles targeting amyloid*

Designing efficient drugs for targeting and interrupting early amyloid self-aggregation is incredible challenging due to the small interaction surface area and lack of higher order structure of amyloid monomers for binding. One suggested way to overcome this is through the use of various nanoparticles (NP). PEGylated long circulating polylactic co-glycolic acid (PLGA) NPs have been shown to capture monomeric and oligomer Aβ in solution and serum, the authors propose that a possible "sink" mechanism may result if used for treating AD [18]. In a similar fashion nanoparticles functionalized with the Aβ ligand, LVFFARK (Figure 1), were shown to protect SH SY5Y cells from amyloid toxicity compared to the peptide alone, which although inhibited fibrillization, also had exhibited strong self-assembly characteristics causing high cytotoxicity. These nanoparticles could potentially act as peripheral sinks, like MAbs and the above mentioned PEGylated NP [67]. As it has been suggested that anionic lipids in the cell membranes cause nucleation sites for amyloid oligomerization, phosphatidic and cardiolipin lipid nanoliposomes were developed which interact with amyloid oligomers and fibrils in solution and serum and could be used to target Aβ [19]. In a more recent study curcumin incorporated into the nanoliposomes were compared to lipid ligand nanoliposomes and shown to be even more effective at inhibiting oligomer and fibril formation [68]. These early molecular studies have led to further progress in delivery and targeting of amyloid which will be discussed later in section 3.

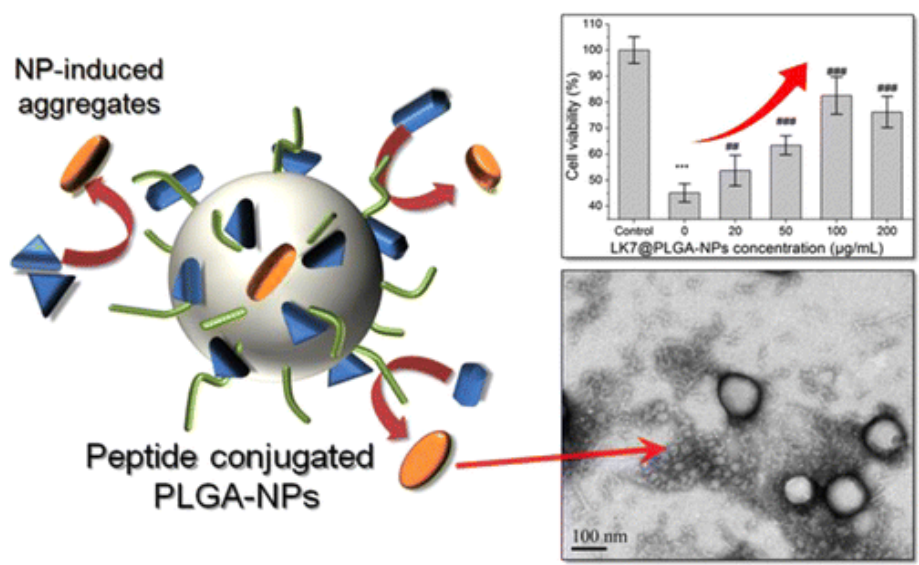

**Figure 1. PLGA nanoparticles modified with Aβ recognition peptide LVFFARK. Shown as green on the cartoon above.** Reprinted with permission from Xiong N, Dong X-Y, Zheng J, et al. (2015) Design of LVFFARK and LVFFARK-functionalized nanoparticles for inhibiting amyloid β-protein fibrillation and cytotoxicity. *ACS applied materials & interfaces* 7: 5650-5662. Copyright © 2015 American Chemical Society.

All of the previous therapeutic strategies, though in principle effective at targeting various aspects of the amyloid cascade, have yet to yield definitive disease-modifying results in human clinical trials causing many to question the amyloid cascade hypothesis [54,56,69,70]. It is likely that amyloid intervention is required prior to the formation of toxic amyloid oligomers [54,56], or that amyloid is not the main cause of neural degradation but a byproduct that causes secondary degeneration [69,70]. To verify the amyloid cascade hypothesis, anti-amyloidosis targeting strategies need to be improved and tested in a relevant population. There are two major challenges that need to be addressed before amyloid targeted therapies can be verified: (1) early diagnostic solutions need to be found so preventative measures can truly be tested, prior to symptom onset and (2) strategies to overcome the restrictive BBB and improve drug and diagnostic probe delivery to the brain are necessary.

## 3.  Drug delivery vehicles and pathways

Aside from the use of NPs as therapeutics against amyloid in AD, nanotechnology has received significant attention for its efficiency through improved surface area in countless other applications. For biomedical purposes and especially delivery across the BBB, there are some necessary and ideal criteria that should be met for safety and efficiency; they should be non-toxic, non-immunogenic, biodegradable, biocompatible, have prolonged blood circulation time, colloidal stability, surface enhancement for BBB permeation, controlled drug release profile and size. The potential efficacy for the delivery of therapeutics across the BBB for the treatment of various diseases using nanoparticles has been well documented using many variations of the same basic structure. These NP drug delivery systems use a basic nanocore structure, such as amphiphilic, polymeric, metallic or hybrid structures. The surface of the NP is important for efficiency and safety, it must be suitably modified for recruitment across the BBB, and finally it will contain the drug target either as a ligand on the surface of, or contained within the NP.

### 3.1. The blood brain barrier (BBB)

The blood brain barrier is a highly selective, regulated and efficient barrier that protects the brain from unwanted molecules and pathogens. It is also the single largest hurdle that needs to be overcome in order to deliver potential therapeutic and diagnostic agents to the brain. The lack of therapeutics for the treatment of Alzheimer's and other brain diseases is not only due to the lack of effective drugs, but due to their inability to cross into the brain from the blood through the endothelium [71]. In the human brain there is approximately 100 billion capillaries and a BBB surface area of 20 m$^2$, as compared to 0.021 m$^2$ for the blood-cerebral spinal fluid barrier [72]. Therefore, most of the entry into the brain is controlled by the BBB, an interface separated by brain endothelial cells on the blood side and astrocytes and pericytes on the brain side [73]. The BBB is comprised of endothelial cells "glued" together with multiple binding proteins (occludins, claudins and junctional adherin molecules) to form tight junctions (TJs) and adherin junctions (AJs). A large portion of brain homeostasis is regulated by influx and efflux at the BBB through these junctions. The BBB has the abilities to prevent entry of and actively remove unwanted molecules from the brain and it regulates the influx of necessary nutrients, signaling molecules and immune cells into the brain. The diverse processes governing the influx of necessary molecules for brain homeostasis provide a variety of options that can be hijacked to improve the delivery of therapeutics and diagnostics into the brain. These processes must be carefully engineered for safety and efficacy due to the fragile nature of the brain [71-73].

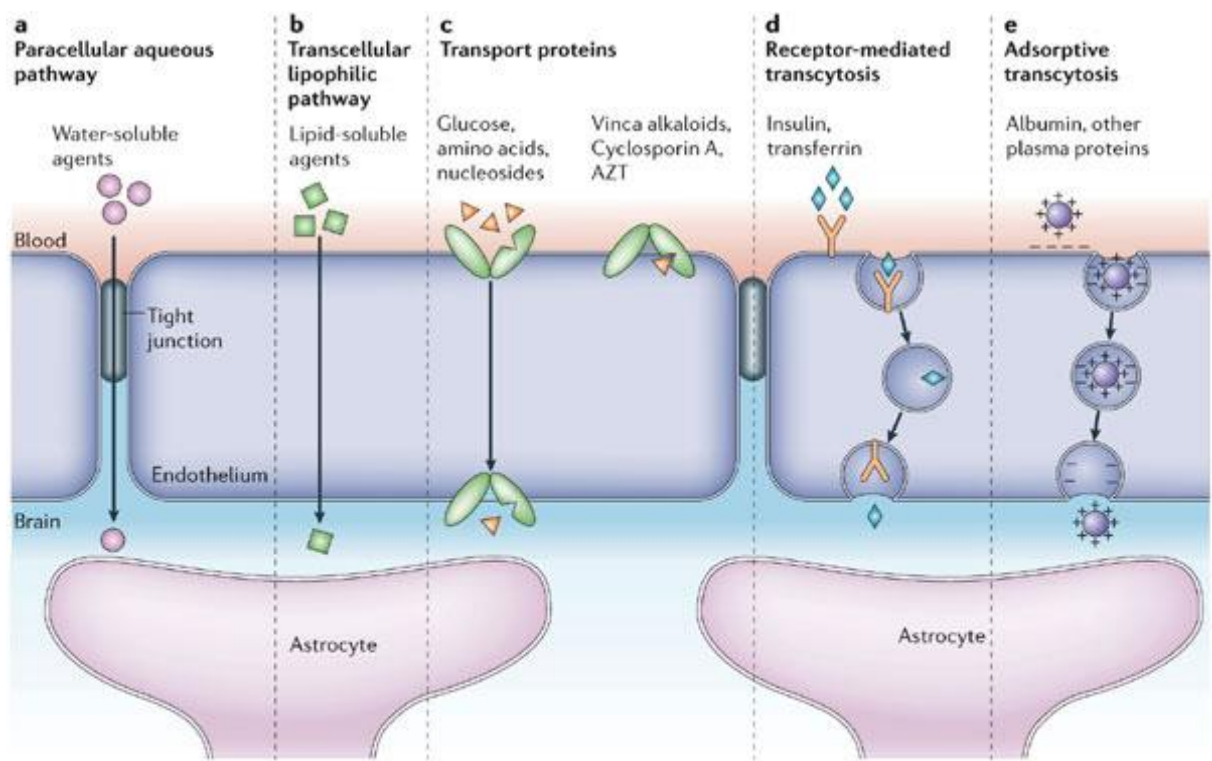

**Figure 2**. **Schematic summary of the possible routes across the blood brain barrier.** Reprinted by permission from Macmillan Publishers Ltd: Nature Reviews Neuroscience: Abbott NJ, Ronnback L, Hansson E (2006) Astrocyte-endothelial interactions at the blood-brain barrier. *Nat Rev Neurosci* 7: 41-53, copyright © 2006.

The essential function of the BBB is to maintain brain homeostasis. It can be characterized through two routes: influx and efflux. The biological properties of the BBB that give it high selectivity are as follows [73]:

a) The BBB has an extremely high resistivity as measured using transendothelial electrical resistance (TEER). This is a measure of the resistance the BBB poses to the flow of charged ions. It has a value between 1500–2000 $\Omega/m^2$ and is a result of the strong barrier posed by TJs between the cells [73]. This resistance prevents access to most charged moieties through TJs so that only small, neutral and water-soluble molecules can pass through these paracellular junctions.

b) The endothelial cells that make up the endothelium/capillary walls of the BBB is supported by astrocytes and pericytes that provide polarization and communication with the endothelial cells to adjust the local properties of the BBB as required, either up-regulating or down-regulating protein and cell membrane composition [74].

c) Various transporter proteins are expressed at the endothelium including glucose carrier (GLUT1) and amino acid carrier (LAT1) for transport of nutrients and small molecules needed for brain metabolism. The transporter protein (cationic-organic transporter protein) is responsible for shuttling small cationic species in and out and are implicated in the transport of current Alzheimer's disease treatments memantine [75] and cholinesterase inhibitors [76] across the BBB.

d) A family of efflux transporters are expressed at the endothelium such as P-glycoprotein (P-gp), multidrug resistance protein (MRP) and receptor for advanced glycation end products (RAGE), along with a class of lipoprotein receptors that are responsible for preventing entry into the brain and actively transporting unwanted material out. RAGE and lipoprotein receptors appear to be mainly responsible for transporting amyloid [77,78].

e) Receptor proteins for influx into the brain are also expressed on the blood side of the brain, for instance: insulin receptors, transferrin receptors (TfR), lipoprotein receptors, diptheria toxin receptors (DTR) and others. These transport larger molecules (mostly peptides, proteins and lipids) into the brain.

f) Strict entry into the brain by immune cells is also permitted by the BBB, for instance perivascular macrophages and monocytes [79].

### 3.2. Transport across the blood brain barrier with respect to AD pathophysiology

The aforementioned features combine to establish a strong physical, transport, metabolic and immunologic barrier. To make use of any of these transport systems for increased uptake of drugs across the BBB, it is important to consider the pathophysiology of the disease. There is no one-size-fits-all drug delivery system for delivery of treatments to all the various neurodegenerative and neurovascular disorders, cancers and infections associated with the brain. This is due to differences in pathophysiology and BBB functionality caused by various diseases and conditions. For example, transporter/receptor proteins can be expressed at different levels in a variety of diseases and TJs in some disorders do not have as high a resistivity as in others. There are many other factors that affect drug delivery in specific conditions. For a more general look at delivery across the blood brain barrier, see Chen & Liu's paper from 2012 [71].

### 3.2.1. Paracellular and Transcellular routes

Only a small set of water soluble molecules and therapeutic drugs are able to pass through TJs (shown in Figure 2a). Utilizing this form of transport into the brain for larger molecules relies on generating transient reversible openings in the TJs between endothelial cells. This can be accomplished through biological, chemical and physical means. Viruses are known to up-regulate cytokines to disrupt the BBB and inflammatory stimuli such as histamine can increase blood-brain permeability. Chemical stimuli, such as cyclodextrin can sequester cholesterol increasing BBB permeation. Physically, it is possible to thermally excite the endothelium using low energy radiation to increase permeation of the BBB (i.e. microwave and ultrasound waves) [71]. These methods are non-specific and will increase the diffusion of most solutes from the blood into the brain.

This route into the brain would not be suitable for a sustained therapeutic treatment over an extended period of time. Chronic conditions like AD would likely require continued treatment from onset until the end of life. Opening the TJs compromises the integrity of the BBB which is already been shown to be impaired [80-83]. The utilization of glucose, insulin receptor distribution and efflux of Aβ are all known to be impaired in AD [84]. Considering this lack of amyloid efflux, it may be beneficial in AD to re-establish a more effective, blood brain barrier rather than compromise it [83-85]. In fact, many researchers consider neurovascular mechanisms of neurodegeneration in Alzheimer's disease as highly relevant and even causative [84,85]. Ultimately, the cost-to-benefit ratio is far too high for disrupting the BBB in AD.

Aside from attempting to pass drugs between cells through the TJs and AJs, molecules can also be passed into the brain through the endothelial cells which make up the BBB. Passive transport occurs for lipophilic molecules and important nutrients through carrier protein pathways. To accommodate larger molecules the cell uses a vesicle mediated transport system called transcytosis. It begins with endocytosis at the luminal (blood) side of the BBB. Here, receptor or electrostatic interactions at the cell membrane trigger vesicle inclusion of the drug or molecule and is then shuttle through the cytoplasm of the brain endothelial cell. Such inclusion of the drug molecule into the vesicle protects it from endogenous enzymes. Following this, the molecule must then undergo exocytosis at the aluminal (brain) side of the BBB. Drugs and molecules using transcellular routes into the brain can do so via various routes, as summarized in Figure 2b–e.

### 3.2.2.    Lipophilic and transport protein pathways (influx/efflux)

Approximately 2% of drugs on the market are able to effectively cross the BBB [42]. Typically, these molecules are small (less than 400 Da), lipophilic, neutral molecules capable of dissolving into the cell membrane, like alcohols and steroidal hormones (see Figure 2b). This is a concentration dependent route into the brain and by itself cannot support delivery of larger macromolecule therapeutics such as peptides and MAbs for targeting amyloid.

As mentioned previously, endothelial cells express transporter molecules for moving glucose, amino acids, nucleotides and other select small molecules into the brain for metabolic needs and signaling pathways (Figure 2c). These transporter proteins are too small to deliver larger macromolecules and nanoparticles into the brain, and their structure and functionality is not well established. To effectively transport drugs using this technique, they must mimic the native nutrient's molecular structure and be about the same size. These transporter proteins move essential nutrients into the brain and trying to exploit them for drug delivery could have negative effects on brain metabolism by competing with essential molecules [71].

Efflux pumps are a family of transporter proteins and receptors expressed in order to move unwanted molecules out of the brain. They are expressed on both sides of the endothelium. They can prevent passage of molecules on the blood side and actively transport them out from the brain side. It is known that amyloid beta binds various efflux and influx pumps [77,78]. For the purpose of delivering and keeping high concentrations of a drug to the brain, one can attempt to design the drug as to not bind P-gp and other efflux pumps. If that is not possible it may be beneficial to utilize efflux pump inhibitors. This will increase concentrations of the drug in the brain by limiting binding events of the drug with the efflux pump. In an AD mouse model it was shown that P-gp deficiency resulted in an increase in amyloid [77]. This suggests that inhibiting P-gp may increase amyloid deposition. Generally, efflux inhibition will interfere with the natural protection offered by the efflux pumps and may not be suitable for treating AD.

### 3.2.3.    Receptor-mediated transcytosis

Receptor-mediated transcytosis (RMT) is a highly specific route across the BBB (Figure 2d). Endothelial cells on the luminal side of the BBB express many cell surface proteins that bind many different ligands including growth factors, hormones, enzymes and other substrates necessary for brain metabolism. Recent advances in molecular biology have made it clear that attaching ligands for receptors that are expressed at the BBB is a feasible way to have selective targeting to the brain. With advances being made in proteomics and genetics it has been possible to more accurately document the pathophysiology of the BBB in various disease states. This helps in narrowing down suitable receptors for targeted delivery to the brain with optimal efficiency for a particular disease. The understanding of expression of proteins at the BBB is not fully developed in the diseased brain, but progress has been made, below are several examples of receptors related to AD pathology:

a) *Transferrin Receptor (TfR)*. Transferrin binds free iron in the blood and the TfR is responsible for transporting these loaded transferrin proteins across various tissue barriers in the body, including the BBB. TfR density in the hippocampus of AD patients is decreased while it is unchanged in cerebral micro-vessels [86], making it a suitable candidate for receptor mediated delivery of drugs in AD. This receptor is a primary target used in RMT nanosystems in the literature. Using transferrin itself as a ligand is not suitable since transferrin receptors are fully saturated by native transferrin in circulation [87]. To overcome this, MAbs and other peptides directed against distinct epitopes of the TfR, which do not compete with binding of native transferrin, have been conjugated to drugs and delivery systems and used to target the brain [87-90]. Of these, the OX26 mouse MAb is the best studied. When conjugated to liposomes and other nanovehicles, it has been shown to improve permeation across the BBB, both *in vitro* and *in vivo* [89]. TfR are heavily expressed at the BBB but in other organs as well including liver, lung and kidney [88]. A rat MAb directed against mouse TfR called R17-217 also undergoes RMT at the BBB (brain uptake at 1.7% dose/g) with very low uptake in kidney or liver when compared to another MAb 8D3, which had a higher brain uptake of 3.1% dose/g [89]. This suggests that the extracellular regions of TfR are different in various cell types and that choosing the correct epitope may allow for targeting to the brain with higher selectivity than other tissue types. This will reduce peripheral tissue uptake and increase brain selective uptake.

b) *Insulin receptors* are important for maintaining glucose homoeostasis in the brain. As Aβ binds readily to insulin receptors [91], glucose utilization is impaired due to the competition of amyloid and insulin with the insulin receptor in AD patients [92]. This receptor may not be an ideal candidate since Alzheimer's patients would not benefit from further glucose impairment.

c) *Lipoprotein receptors* are a broad class of receptor proteins that are responsible for scavenging and signaling in the brain. It is known that down-regulation of low density lipoprotein receptor related protein (LPR-1) decreased clearance of Aβ molecules from the brain. Competition with an external ligand for drug delivery may impede the clearance of Aβ from the brain, which is presumably detrimental in AD. On the flip side, another low density lipoprotein, the ApoE receptor is responsible for shuttling cholesterol into the brain and thus has been suggested as a target for improved BBB permeation [93,94].

d) *Diphtheria toxin receptor (DTR)* may be a good candidate for use in AD BBB delivery since it has no ligands necessary for brain homoeostasis [95]. DTR is also known to be up-regulated in conditions of inflammation in the brain which is associated with AD [95]. This may be a useful target for both imaging and therapeutic delivery in AD. Due to the toxicity of the diphtheria toxin, it is not a suitable ligand for drug delivery. A non-toxic mutant CRM197 has been explored for targeting the DTR to improve drug delivery. It has been used as a component of vaccines to increase immune response since the 1980s and therefore has a track record of being safe [71]. The transport capacity using CRM197 was tested using a protein tracer across an *in vitro* model and *in vivo* using guinea pigs. They found that transcytosis occurred at a slow rate after an initial delay. It is suggested that antibody to CRM197 may be present in serum; if so it would neutralize its ability to bind to the DTR [95]. That being said, it may be possible to target other epitopes of DTR using shorter non-immunological peptide sequences.

e) *Nicotinic acetylcholine receptors* are the target of rabies virus glycoprotein (RVG), a 39 amino acid peptide. RVG binds the alpha-7 subunit, which is expressed in neurons and endothelial cells; as such it has been widely explored as a brain targeting ligand especially in RNA interfering strategies [96,97].

Receptor-mediated transcytosis has great potential for brain specific targeting of drug delivery systems due to its specificity and increased BBB permeation. However, high affinity does not guarantee highly efficient transcytosis. Yu et al. showed that a high affinity anti-TfR antibody had lower levels of transcytosis than the lower affinity versions [66]. This implies that there is an ideal affinity that will achieve a balance between selectivity and capacity. In a study by Couch et al., bispecific MAbs directed at TfRs and BACE1 with reduced affinity showed improved brain uptake and safety while reducing peripheral exposure [98]. This same group was able to use imaging and colocalization experiments to determine the fate of MAbs with various affinities, they found that high affinity MAbs were trafficked to lysosomes for degradation and also significantly decreased TfR activity reducing further uptake of new doses [99]. TfR/BACE1 bispecific antibodies showed increased BBB penetration and reductions in brain amyloid compared to BACE1 MAb alone in non-human primates; as well they showed that there is an ideal TfR affinity that is neither too high nor too low [100]. Similarly, MAb directed against Aβ have been conjugated with single Fab MAb fragments that target TfRs and have demonstrated much improved BBB permeation. This monovalent targeting of TfRs caused increases in brain deposition 55-fold while in contrast using bivalent (two Fab MAb fragments), which increases affinity, resulted in lysosome trafficking and reduced BBB permeation [101].

If a receptor is necessary for transporting essential nutrients or regulating brain homoeostasis, it may not be a good target for RMT as competition in the compromised disease state may inflict more harm than the drug will benefit. Therefore it is recommended that the ligand bind a different portion of the receptor than the native substrate. A great challenge in determining the ideal receptor targeting ligand is the lack of consistency in the system being tested and the model used for assessing BBB permeation. In a step forward, five different RMT targeting ligands were tested in parallel using identical liposomes *in vivo*. In comparing these five ligands only one MAb, R17-217 had sufficient increases in BBB permeation at all of the time points while the other receptor ligands did not greatly affect liposome delivery [102]. The results are summarized in Figure 3.

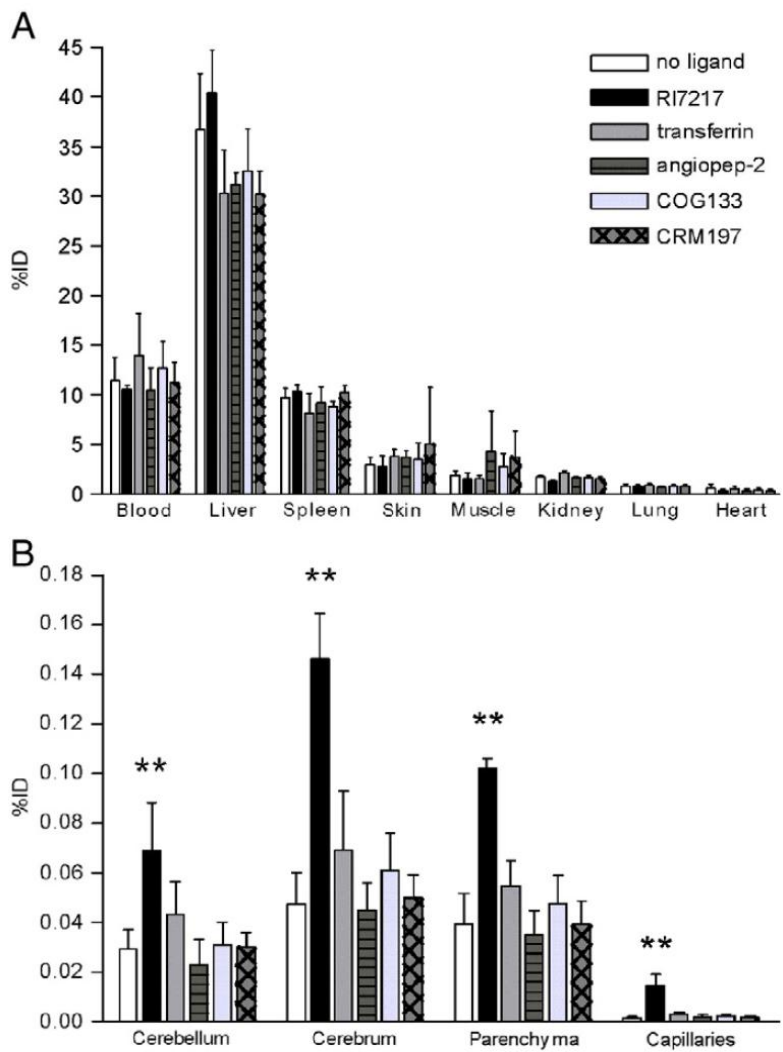

**Figure 3. Distribution of liposomes with various targeting ligands for RMT A) in the body and B) in the brain at 12 hours.** ID stands for injected dose. Reprinted from Journal of Controlled Release, 150/1, van Rooy I, Mastrobattista E, Storm G, et al. Comparison of five different targeting ligands to enhance accumulation of liposomes into the brain, 30-36, Copyright 2011 with permission from Elsevier.

Nanoparticle delivery systems using receptor mediated transcytosis

In a study performed by Prades et al., they used a ligand for TfR to improve the delivery of gold nanoparticles across the BBB. The gold nanoparticles (AuNP) were conjugated to a β-sheet breaker peptide six amino acids long (CLPFFD) [90]. These AuNP-CLPFFD are able to recognize amyloid and after irradiation with weak microwaves break up amyloid deposits as a form of molecular surgery. These AuNP-peptide nanocomposites are unable to cross the BBB. To improve permeation through the BBB they conjugated a peptide ligand for TfR. This short 12 amino acid sequence (THRPPMWSPVWP) may be advantageous over MAb as it has high specificity and small size (necessary to prevent steric hindrance of the small amyloid recognition peptide). The nano-vehicles and controls are under 20 nm, which is favourable, unlike their highly negative ζ-potential that contributes to reduced BBB permeation. To assess BBB permeation, a model comprising bovine-brain endothelial cells (BBEC) on top of a filter, with rat astrocytes grown on the bottom was used (Figure 4). This creates a supported and polarized model BBB with the top compartment acting as the luminal (blood) side and the bottom acting as the aluminal (brain) side of the BBB. The BBB permeation of the nanosystem can then be tested by introducing them in the top compartment and measuring the amount in the bottom over time. The TEER of this model is about an order of magnitude less than what the human BBB is expected to be in vivo, this is indicative of adequate TJ formation for this type of *in vitro* assessment. By conjugating the nanocarrier with THR (the ligand for TfR), an increase in BBB permeation was observed, by a factor of almost 1200 and a factor of 12 in the presence fetal bovine serum. Passive diffusion is not possible as gold nanoparticles were unable to permeate in a parallel artificial membrane permeability assay, PAMPA for short.

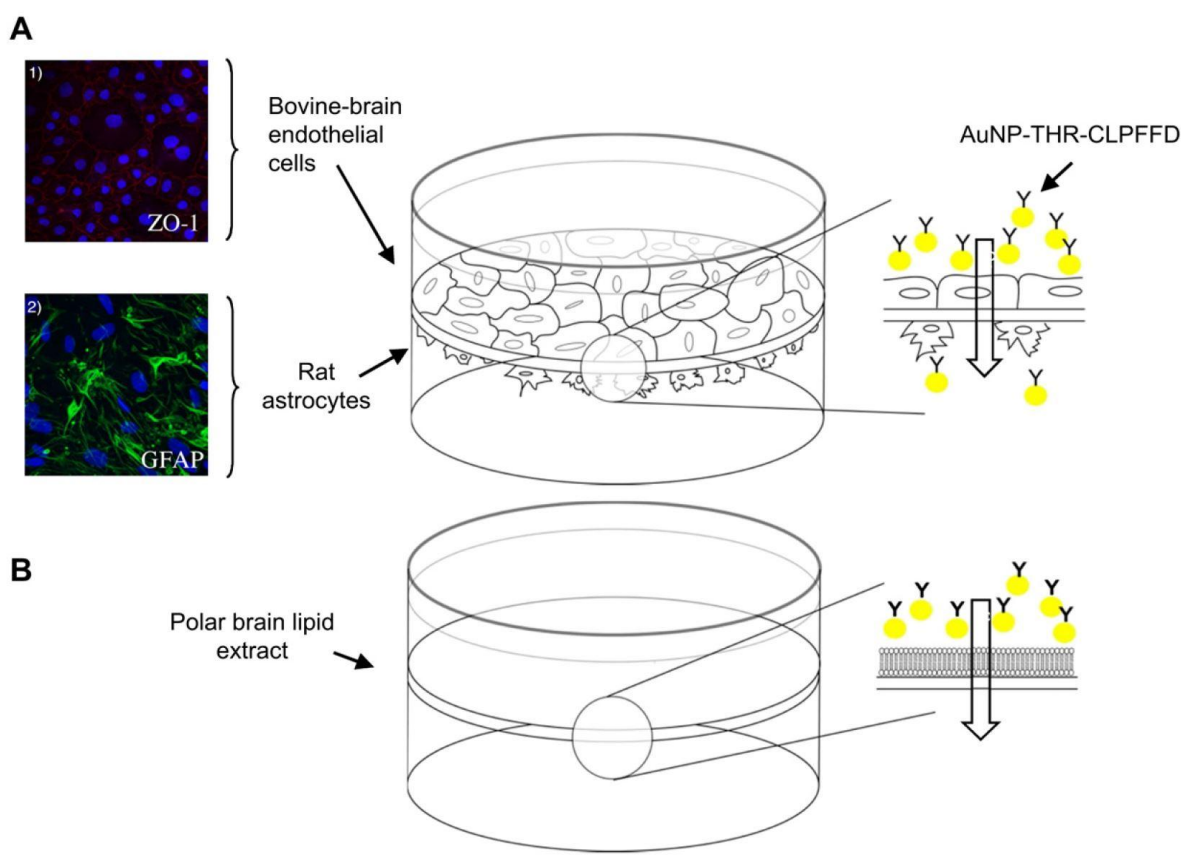

**Figure 4. A)** *In vitro* **model to test the efficacy of AuNP-THR-CLPFFD and control nanocarrier's ability to permeate the BBB. B) PAMPA to test BBB permeation through passive diffusion.** Reprinted from Biomaterials, 33/29, Prades R, Guerrero S, Araya E, et al.. Delivery of gold nanoparticles to the brain by conjugation with a peptide that recognizes the transferrin receptor, 7194-7205, Copyright 2012 with permission from Elsevier.

Lastly, *in vivo* assessments of BBB permeation on rats was performed. In this portion of the study the rats were given a peritoneal injection of the THR-CLPFFD conjugated or control AuNPs.

After, the animals were analyzed postmortem at 30 minute, 1 hour and 2 hour time intervals. The gold content of the brain and peripheral organs was assessed using neutron activation. They found high levels of gold in the liver and spleen, as to be expected (38 % and 18% of ID, respectively); low levels of gold accumulation in the brain (0.07% of the ID) was sustained for at least an hour. This was twice as high as the control AuNP accumulation after 1 hour. It was shown that an increase in the uptake of nanoparticles can be achieved using TfR ligands, despite unfavorable electrostatic surface potential. It's safe to suggest that a system with favorable energetics would have far improved uptake at the BBB, compounding and complementing RMT.

In a recent study, Zhang et al. produced a dually functional nanosystem for delivery across the blood brain barrier and targeting of Aβ [103]. They produced two peptide ligands, one for TfR and one for Aβ, using phage display. These peptide ligands were then conjugated to the surface of a polyethylene glycol-poly lactic acid polymer (PEG-PLA) nanoparticle. These nanoparticle systems were approximately 100 nm in size, with Zeta (ζ) potential around −22 mV. MTT assays showed negligible cytotoxicity while *in vivo* BBB permeation studies involving coumarin-6 dyes under confocal microscope showed an increase in deposition in the brain [103]. *In vivo* studies focused on blood brain barrier permeation for the purpose of evaluating depositions in the brain. Since no drug or diagnostic payload was delivered, it was not necessary to monitor long term cognitive effects. This study further highlights functionalized nanoparticles with peptides directed at TfR as an effective strategy to promote increased BBB permeation in Aβ challenged brains. The authors suggest that such a system may be useful to deliver a diagnostic or therapeutic. In a follow up study this dually functional nanosystem was used to deliver a generic beta sheet breaker peptide, called H102, to mice that had received intracranial injections of Aβ. A "sham" group received intracranial injections of isotonic saline and was used as control [104]. The H102 bi-functional NPs exhibited neuroprotective effects with two key biochemical indexes being restored to sham levels and histology revealing reductions in pathology and cell loss in the hippocampus, while improved spatial learning and memory was seen in the Morris water maze.

Based on earlier work done with lipid nanoparticles containing phosphatidic acid for targeting amyloid [19], similar nanoliposomes were modified with lipoprotein receptor ligands (ApoE peptides) to improve BBB permeation [94]. Later, this same group, again using phosphatidic acid NPs to target amyloid, compared the effects of various linkages of MAbs directed at transferrin receptors for improving BBB penetration. They showed that covalent linkage of anti-TfR MAb R17-217, using Mal-PEG, to the liposome improves the ability of the liposome to breach the BBB as compared to a biotin-streptavidin conjugation [105]. A different group used dually decorated nanoliposomes with Aβ MAb and OX26 anti-TfR MAb using biotin streptavidin conjugation for improved delivery across the blood brain barrier [106] (Figure 5 below). Subsequently, this group tested anti-TfR MAb and an ApoE derived peptide as dual targeting ligands for improved absorption *in vitro*, with a BBB model and *in vivo* animal studies. They found that both receptor ligands improved uptake at the endothelium and had a compounding effect *in vitro* but not *in vivo*. These conflicting reports were reconciled when serum was added to the cellular BBB model, at which point they found ApoE was ineffective at improving uptake in the endothelial cells [93]. This may be due to competitive processes between serum proteins and target receptor, or because serum proteins directly quench ApoE peptides. These reports suggest that covalent conjugation of the targeting ligands are superior to biotin-streptavidin systems and that transferrin receptor ligands may be an ideal choice of targeting ligand over ApoE peptides.

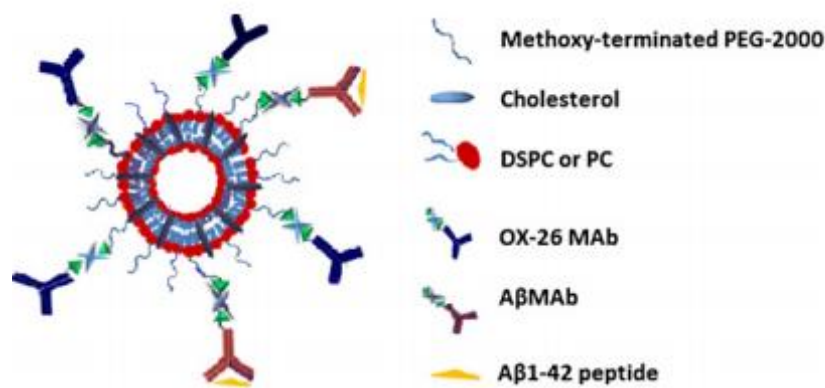

**Figure 5**. **Dually decorated nanoliposomes targeting both TfR and Aβ with MAb.** Reprinted from European Journal of Pharmaceutics and Biopharmaceutics, 81/1, Markoutsa E, Papadia K, Clemente C, et al. Anti-Abeta-MAb and dually decorated nanoliposomes: effect of Abeta1-42 peptides on interaction with hCMEC/D3 cells, 49-56, Copyright 2012 with permission from Elsevier.

3.2.4. Absorptive-mediate transcytosis

Native positively charged proteins and macromolecules such as human serum albumin are capable of transcytosis across the BBB through electrostatic interactions with the negatively charged micro domains in the endothelial cell membrane; this process is called absorptive-mediated transcytosis or AMT (Figure 2e). By enhancing the surface of nanoparticle drug delivery systems with positively charged moieties, AMT has the potential to greatly improve the permeation of many drugs at the BBB. AMT begins with a non-specific electrostatic interaction with the endothelial cell membrane and as such has a high capacity for improving BBB permeation [107]. As a result of the low specificity of AMT these drug delivery system will have increased uptake in other tissues in the body, such as the liver and kidney. The increase in permeation into other organ systems leads to less drug delivered to the brain. This challenge is not insurmountable as it is known that polyethylene glycol (PEG) coated nanoparticles have decreased liver uptake and increased blood circulation time [18]. Moreover, by combining AMT favorable surface enhancement with RMT, it may be possible to have efficient, high capacity, and high specificity drug delivery across the BBB for the treatment of various disorders including AD.

Several studies have shown that molecules with positive moieties and positive ζ potentials can increase permeation through the blood brain barrier [107]. Protamine [108], cationic proteins [109], and a class of cell-penetrating peptides (CPPs) [110] have been used to improve AMT. CPPs are an especially versatile class of short (less than 30 amino acids long) peptides capable of improving penetration into cells. The exact mechanisms governing CPP are currently in debate and, in fact, many mechanisms may be acting in parallel; evidence suggests cationic CPPs may involve endocytosis pathways as in AMT [110]. A CPP derived from the HIV trans-activator of transcription, TAT for short, was retro inversed and conjugated to the previously mentioned RI-OR2 anti-aggregation inhibitor with remarkable results in an *in vivo* transgenic mouse model [65]. The TAT/RI-OR2 peptide was able to cross the blood brain barrier and bind plaques, after 21 days of intraperitoneal injections dual transgenic mice had 25% reduction in cerebral cortex Aβ, 32% reduction in plaque count, 44% reduction in activated microglial cells, 25% reduction in oxidative damage and 210% increased neuron count in the dentate gyrus. These results indicate that oxidative damage, reduced neurogenesis and inflammation are results of Aβ aggregation and can be reduced with the use of peptide inhibitors provided they can be delivered to the brain.

Aside from CPPs, nanostructures which contain a positive charge hold great promise as either modifications to surface enhanced drug delivery systems or as the core nanomaterial themselves. Cationic polymers, typically featuring amine groups, have been shown to have to increase permeation at the BBB. This is due to the protonation of these groups in neutral to acidic conditions such as physiological pH [71,111]. Polycationic polymers are known to be cytotoxic, are not typically suitable for clinical applications, and their mechanism is not well understood, though both necrotic and apoptotic factors are involved [112]. One natural cationic polymer stands out, several studies have shown that chitosan conjugated nanoparticles can increase uptake at the BBB and do so safely [111,113]. Chitosan is a natural polymeric molecule comprised of randomly distributed polysaccharides containing amine groups and has been of great interest as a nanoparticle drug carrier in many diseases including AD [114]. These amine groups become positively protonated in physiological pH this generates a favorable surface potential for endosome forming cell membrane interactions. Many groups have utilized chitosan as a safe and effective polymeric delivery vehicle [111,113,115,116].

Nanosystems utilizing absorptive-mediated transcytosis to cross the BBB

One model nano-vehicle consists of a polylactic-co-glycolic acid (PLGA) polymer nanocore, surface enhanced with chitosan for improved electrostatic surface potential to facilitate AMT and anti-amyloid IgG4.1 antibody for targeting of amyloid deposits [111]. PLGA is a biodegradable, biocompatible, non-toxic, non-immunogenic copolymer of lactic and glycolic acid. The ratio of lactic to glycolic acid in this co-polymer change its rate of degradation and in a drug delivery system will allow for a tunable drug release profile. A higher ratio of glycolide to lactide lowers the degradation time of PLGA, with the exception of a 50:50 ratio which leads to fastest degradation [111]. The surface enhancing molecule chitosan is a naturally occurring polysaccharide; it is made of randomly distributed β-linked ᴅ-glucosamine (deacetylated unit) and *N*-acetyl-ᴅ-glucosamine (acetylated unit). Since chitosan contains amine groups, it becomes protonated in acidic to neutral conditions. Its ζ potential is around 20 mV at a pH of 5 and around 10 mV at a pH of 7 (physiological conditions). This positive ζ potential contributes to interactions at the endothelial cell membrane and also improves its colloidal stability. Chitosan's properties can be modified in many ways and also satisfies the conditions required by nanoparticle drug delivery systems with regards to biocompatibility, biodegradation, toxicity and immunogenicity. Chitosan was also shown to be an effective cryo-protectant, which is important for long-term stability during storage of drugs and therapeutics.

Using a polarized Madin-Darby canine kidney (MDCK) cell monolayer model to assess BBB, an increase in the uptake of immuno-nanovehicles conjugated with chitosan verses the control was found [111]. This shows that chitosan can induce AMT at the BBB. Moreover, cells challenged with Aβ had a very large increase in the uptake of immuno-nanovehicles, whether this is due to targeting of Aβ or because Aβ reduces the integrity of the BBB remains to be seen. It is likely that both may factor into the total BBB permeation. These results were visually demonstrated using confocal laser microscopy shown in Figure 6. The utility of using chitosan nanostructures for improved AMT across the BBB, for targeting Aβ, has been shown. However, verification of its ability to sufficiently cross the BBB in vivo is necessary. Since this nanovehicle does not have selectivity for brain and its surface potential will likely result in low blood circulation time and therefore a reduced uptake at the BBB due to clearance of the nanovehicle in the liver, kidneys and lungs. Possible improvements would include adding targeting ligands for the brain and PEG conjugation to reduce liver uptake. This PEG conjugation, however, is associated with a lowering of ζ potential. More work needs to be done to verify whether or not this system would achieve sufficient concentrations *in vivo*, and characterize the deposition in other body tissues.

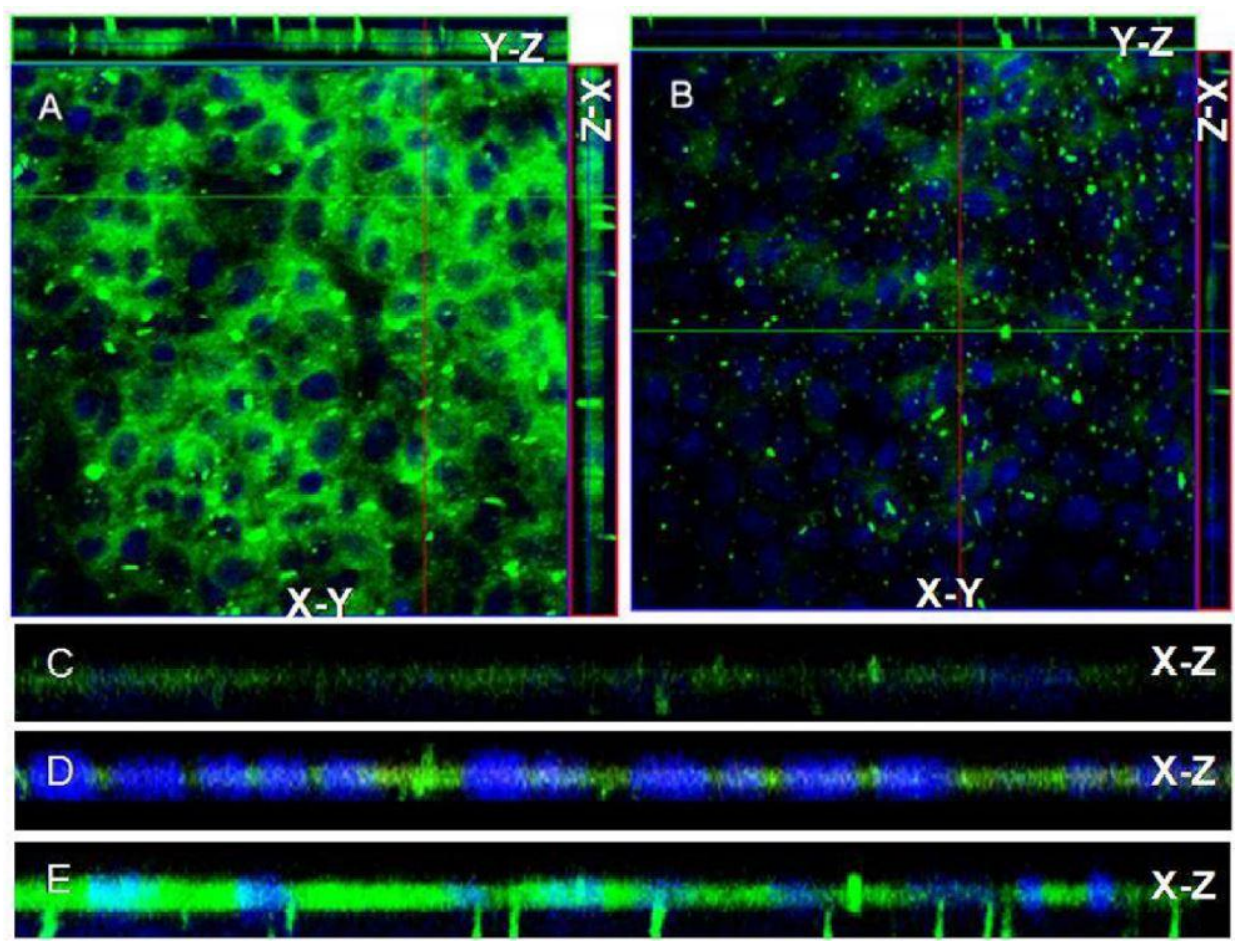

**Figure 6. Laser confocal microscopy images of MDCK cell monolayer under various test conditions.** A) Uptake of CPLGA immuno-nanovehicles (IgG4.1 present) encapsulated with 6-coumarin (green dye). B) Control immune-nanovehicles (PLGA-IgG4.1, no chitosan present), notice the lack of 6-coumarin, implies low uptake. C) 6-coumarin chitosan conjugated PLGA nanoparticle (no IgG4.1 present). D) Immuno-nanovehicles in MDCK monolayer, note moderate absorption. E) Immuno-nanovehicles in MDCK monolayer challenged with Aβ, note large amounts of endocytosis and transcytosis (columns of bright green). Reprinted from Nanomedicine: Nanotechnology, Biology and Medicine, 8/2, Jaruszewski KM, Ramakrishnan S, Poduslo JF, et al. Chitosan enhances the stability and targeting of immuno-nanovehicles to cerebro-vascular deposits of Alzheimer's disease amyloid protein, 250-260, Copyright 2012 with permission from Elsevier.

Polymer-based nanoparticle drug delivery systems are readily modified. Chitosan PEG copolymers represent a balance between favorable blood circulation time and ζ potential, especially for the delivery of gene therapies and negatively charged peptides. Moreover, there are safety concerns associated with cationic polymer nanoparticles as they may cause disruption of the cell membrane. The use of PEG to quench some of this positive charge may improve safety concerns. Malhotra et al. were able to conjugate up to 7.5% of the amine groups of chitosan with PEG, forming nanoparticles less than 100 nm [113]. Afterwards, they used these PEG molecules as linkers for a cell-penetrating peptide to increase AMT. They showed this significantly improved uptake of siRNAs delivered to neuroblastomas *in vitro*.

## 4. Future directions

The delivery of therapeutics across the BBB for the treatment of Alzheimer's disease can be improved using nanoparticles. The high capacity utility of AMT and the specificity of RMT are both highly desirable to efficiently deliver compounds to the brain. The improved delivery of drugs across the BBB has been shown through the use of many different systems and tested using a wide array of methodologies, *in vitro* and *in vivo*. All of these studies show promise, however the lack of consistent approaches has made it difficult to track down any one best method for a given disorder. That being said, it is still possible to use these preliminary studies to think critically about novel systems that will have improved efficacy over previous ones. These nanoparticle drug delivery systems use a basic nanocore structure (amphiphilic, polymeric, or metallic), a targeting ligand for Aβ and some other surface enhancement for BBB permeation, either AMT or RMT. Utilizing both AMT and RMT to improve the surface characteristics of the nanoparticle may improve the efficiency of BBB permeation for delivering therapeutics.

Liposomes can go through bi-ligand modification for delivery to neurons that utilizes both TfR targeting with native transferrin (for RMT) to improve specificity and a poly-L-arginine CPP to improve capacity through AMT [116]. In this particular example, the β-galactosidase reporter gene is delivered *in vivo* to adult rats. The liposomes were conjugated with poly-L-arginine, which is protonated in most acidic, neutral, and basic conditions. This resulted in a final ζ potential around 10 mV for bi-ligand nanovehicle. The TfR ligand used was found to reduce the zeta-potential by approximately 10 mV. They were able to achieve a deposition of 4% of the injected dose per gram of tissue at a 24 hour time point, which was twice the injected dose (2% ID/mg tissue) as seen in the transferrin modified liposome alone. Less than 1% was seen in the control liposome. A detailed analysis of the deposition of bi-ligand liposomes in the various organs was performed, with considerable uptake in the liver, spleen, kidney, lungs and heart. This is not surprising as transferrin receptors are expressed in all these tissue types. This system may be improved by utilizing a peptide or MAb ligand more specific to brain TfR which does not complete with native transferrin homeostasis, such as R17-217 MAb.

Another system that combines RMT and AMT pathways for improved uptake at the BBB is an RVG-peptide-linked trimethylated chitosan for delivery of siRNA against BACE1 transcripts to the brain [115]. RVG is known to bind nicotinic acetylcholine receptor subunit alpha-7 expressed at the luminal side of the endothelium as well as on neurons. This improves uptake at the BBB and at the neuronal cell membrane. The nanocore was built out of trimethylated chitosan combined to form a copolymer with PEG, as similar to what was seen previously. The final siRNA delivery vehicle had a $\zeta$ potential that decreased with the ratio of positively charged amino groups on chitosan to negatively charged siRNA. At a ratio of 96:1 the zeta potential was $9 \pm 2.5$ mV for good AMT capacity. This system benefits from the serum stability, and thus longevity, afforded by PEG and has the targeting capability of RVG. *In vitro* gene silencing showed a 50% reduction in BACE1 levels while *in vivo* imaging showed significant deposition in the brain compared to background levels found in the control, simultaneously deposition in the kidney and liver decreased slightly. There was no deposition found in the heart or spleen and hepatotoxicity was not observed. This study shows promise for the development of more efficient drug delivery vehicles. Long-term studies involving behavior and effect on cognition in AD mouse models are necessary to verify the safety and efficacy of this BACE1 delivery system.

## 5. Conclusion

Over 100 years have passed since the first characterization of AD and although significant progress in the molecular mechanisms has been seen, there is still much that needs to be clarified. Of particular importance is the verification of the amyloid cascade hypothesis. It has already been determined that A$\beta$ directed therapeutics are not sufficient to treat mild to moderate AD and a shift in paradigms to preventative clinical trials has begun in the last couple of years. It is important to note that many current strategies being explored in preventive clinical trials were not designed to intervene in a preventative manner and that other therapeutics may offer significant advantages. Moreover, a null result may still be seen if efficient delivery of therapeutics to the brain cannot be achieved. Regardless of the success or failure of the amyloid cacscade hypothesis, better drug delivery systems for improving delivery of pharmaceuticals to the brain are still a priority and lessons learned from targeting A$\beta$ can be applied to the delivery of other drugs for treating AD should they prove more relevant in the future.

## Acknowledgments

The authors thank Dr. E. Drolle and Dr. R. Henderson for helpful discussion and Dr. F. Hane for critical reading of the final manuscript. This work has been funded by Natural Science and Engineering Council of Canada (NSERC) and University of Waterloo Chronic Disease Prevention Initiative (CDPI) grants to ZL.

## Conflict of interest

Authors declare no conflict of interest.

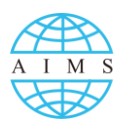 AIMS Press